\input{aipcheck.tex}

\newcommand\sps{\space\space\space\space}
 \def\selectedoptions{final}

\documentclass[
   \selectedoptions
  ]
  {aipproc}

\def\selectedlayoutstyle{8x11double}
\layoutstyle\selectedlayoutstyle
\SetInternalRegister\hbadness{8000} 

\newcommand\doingARLO[2][]{%
  \ifx\mmref\undefined #1\else #2\fi
}

\begin{document}

\title{Chiral Symmetry and Neutrino Pion Production off the Nucleon }


\author{E. Hern\'andez}{
  address={Grupo de F\'\i sica Nuclear,
Departamento de F\'\i sica Fundamental e IUFFyM,\\ 
Universidad de Salamanca, E-37008 Salamanca, Spain},
  email={gajatee@usal.es},}

\iftrue
\author{J.~Nieves}{
  address={Departamento de
 F{\'\i}sica At\'omica, Molecular y Nuclear,\\ Universidad de Granada,
 E-18071 Granada, Spain},
  email={jmnieves@ugr.es},
}

\author{M.~Valverde}{
  address={Departamento de
 F{\'\i}sica At\'omica, Molecular y Nuclear,\\ Universidad de Granada,
 E-18071 Granada, Spain},
  email={mvalverd@ugr.es}}
\fi

\copyrightyear  {2001}

\begin{abstract}
  The neutrino pion production off the nucleon is traditionally
  described in the literature by means of the weak excitation of the
  $\Delta(1232)$ resonance and its subsequent decay into $N\pi$. Here,
  we present results from a model that includes also some background
  terms required by chiral symmetry.  We show that the contribution of
  these terms is sizeable and leads to significant effects in total
  and partially integrated pion production cross sections at
  intermediate energies of interest for neutrino oscillation
  experiments. Finally, we discuss parity-violating contributions to
  the pion angular differential cross section induced by the
  interference of these non-resonant terms with the $\Delta$ piece.
\end{abstract}

\date{\today}

\maketitle

\section{Introduction}

The interest of studying pion production off the nucleon or off some
nuclei like oxygen or carbon is twofold. First, it helps to unravel
some aspects of the hadronic structure which are not accessible to
electromagnetic probes. Second, these processes play an important role
in the analysis of the present neutrino oscillation experiments, where
they constitute a major source of uncertainty in the identification of
electron and muon events. In this talk, we focus on the neutrino and
antineutrino pion production off the nucleon driven both by Charged
and Neutral Currents (CC and NC) at intermediate energies. The model
presented here will allow us to extend the results of
Refs.~\cite{ncjuan} for NC and Ref.~\cite{ccjuan1} for CC driven
neutrino-nucleus reactions in the quasielastic region, to higher
nuclear excitation energies up to the $\Delta(1232)-$resonance region.

There have been several studies of the weak pion production off the
nucleon at intermediate energies~\cite{LS72}--\cite{Paschos}. Most of
them are based on the dominance of the weak excitation of the
$\Delta(1232)$ resonance and its subsequent decay into $N\pi$,
neglecting thus any contribution arising from non-resonant mechanisms.
Here in addition, we also consider some background terms required by
the pattern of spontaneous chiral symmetry breaking of QCD. Their
contributions\footnote{Some non-resonant terms were also studied in
  Refs.~\cite{FN} and~\cite{SUL03}. In the latter reference, the
  chiral counting was broken to account explicitly for $\rho$ and
  $\omega$ exchanges in the $t-$channel, while the first work is not
  consistent with the chiral counting either and it uses a rather
  small axial mass ($\approx 0.65$ GeV), as well.} are sizeable and
lead to significant effects in total and partially integrated pion
production cross sections even at the $\Delta(1232)-$resonance peak,
and they are dominant near pion threshold. As a consequence, when we
compare the predictions of the model with the ANL bubble chamber cross
section data~\cite{anl}, we need to use an axial $\Delta N W$ strength
substantially smaller than that traditionally assumed in the
literature and deduced from the off diagonal Goldberger-Treiman
relation (GTR). Corrections to GTR turn out to be smaller if the BNL
data~\cite{bnlviejo} were considered. Some more details and extended
results can be found in Ref.~\cite{prd}.

\section{The $W^+N \to N'\pi$ reaction }

\begin{figure}
\includegraphics[height=.28\textheight]{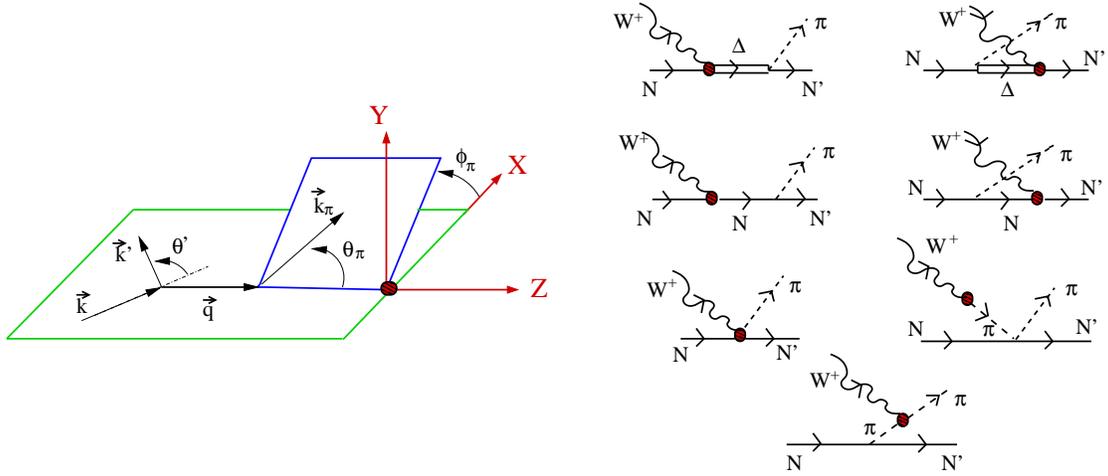}
  \caption{Left: Definition of the different kinematical variables
used through this work. Right: Model for the $W^+ N\to N^\prime\pi$
  reaction. It consists of seven diagrams: Direct and crossed
  $\Delta(1232)-$ (first row) and nucleon (second row) pole terms,
  contact and pion pole contribution (third row) and finally the
  pion-in-flight term.  The circle in the diagrams stands for the weak
  transition vertex.}
\label{fig:diagramas}
\end{figure}

We will focus on the  $\nu_l (k) +\, N(p) \to l^-
(k^\prime) + N(p^\prime) +\, \pi(k_\pi)$ reaction, though the
generalization to antineutrino induced reactions and/or NC
driven processes is straightforward~\cite{prd}. The
unpolarized LAB fifth differential cross section 
(kinematics is sketched in Fig~\ref{fig:diagramas}) is determined by
the contraction of the leptonic ($L$) and hadronic ($W$) tensors, 
\begin{eqnarray}
&&\frac{d^{\,5}\sigma_{\nu_l
    l}}{d\Omega(\hat{k^\prime})dE^\prime d\Omega(\hat{k}_\pi) } 
=\nonumber\\
&&   
 \frac{|\vec{k}^\prime|}{|\vec{k}~|}\frac{G^2}{4\pi^2}
    \int_0^{+\infty}\frac{d|\vec{k}_\pi| |\vec{k}_\pi|^2}{E_\pi}
    L_{\mu\sigma}^{(\nu)}\left(W^{\mu\sigma}_{{\rm CC}
    \pi}\right)^{(\nu)}\label{eq:sec}
\end{eqnarray}
with $L_{\mu\sigma}^{(\nu)} = k^\prime_\mu k_\sigma +k^\prime_\sigma
k_\mu - k\cdot k^\prime g_{\mu\sigma} + {\rm i}
\epsilon_{\mu\sigma\alpha\beta}k^{\prime\alpha}k^\beta $, while the
hadronic tensor is written in terms of the matrix elements of the
quark CC $j^\mu_{\rm cc+}$,
\begin{eqnarray}
  (W^{\mu\sigma}_{{\rm CC} \pi})^{(\nu)} &=&
  \overline{\sum_{\rm spins}}
  \int\frac{d^3p^\prime}{(2\pi)^3} \frac{
    \delta^4(p^\prime+k_\pi-q-p)}{8ME^\prime_N}\nonumber \\ 
 &\hspace{-1cm}\times& \hspace{-0.75cm}\langle N^\prime \pi |
  j^\mu_{\rm cc+}(0) | N \rangle \langle N^\prime \pi | j^\sigma_{\rm
    cc+}(0) | N \rangle^*
\label{eq:wmunu}
\end{eqnarray}
with $q_\mu =k_\mu-k^\prime_\mu$, $M$ the nucleon mass. Lorentz invariance
restricts the dependence of the cross section on $\phi_\pi$ (pion
azimuthal angle),
\begin{eqnarray}
\frac{d^{\,5}\sigma_{\nu_l
    l}}{d\Omega(\hat{k^\prime})dE^\prime d\Omega(\hat{k}_\pi) } &=&
\frac{|\vec{k}^\prime|}{|\vec{k}~|}\frac{G^2}{4\pi^2}
     \Big \{ A +
    B \cos\phi_\pi  \nonumber \\
 &\hspace{-2.2cm}+& \hspace{-1.35cm} C \cos 2\phi_\pi+
    D \sin\phi_\pi + E \sin 2\phi_\pi \Big\} \label{eq:phipi}
\end{eqnarray}
with $A,B,C,D$ and $E$  real, structure functions,
which depend on $q^2,\, p\cdot q$, $k_\pi\cdot q$ and $k_\pi\cdot p $. 
To compute the background contributions to $\langle N^\prime \pi |
j^\mu_{\rm cc\pm}(0) | N \rangle$, we start from a chiral SU(2)
non-linear $\sigma$ model involving pions and nucleons.  We derive the
corresponding Noether's vector and axial currents and those determine,
up to some form-factors (constrained by CVC and PCAC and the
experimental data on the $q^2$ dependence of the $WNN$ vertex), the
contribution of the chiral non-resonant terms. To include the $\Delta$
resonance, we parametrize the $W^+ N \to\Delta$ hadron matrix element
as in~\cite{LS72} with the set of form-factors used in
~\cite{Paschos}.  In particular, for the dominant $C_5^A(q^2)$ axial
form--factor, we initially take
\begin{equation}
C_5^A(q^2) = \frac{1.2}{(1-q^2/M^2_{A\Delta})^2}\times
\frac{1}{1-\frac{q^2}{3M_{A\Delta}^2}}  \label{eq:ca5old}
\end{equation}
with $M_{A\Delta}=1.05 $ GeV and $C_5^A(0)$ is fix to the prediction
of the off-diagonal GTR.  The model consists of seven diagrams
(right panel of Fig.~\ref{fig:diagramas}) and it is an extension of
that developed in Ref.~\cite{GNO97} for the $ e N \to e' N \pi$
reaction.
\section{Discussion}

\begin{figure}
\includegraphics[height=.28\textwidth]{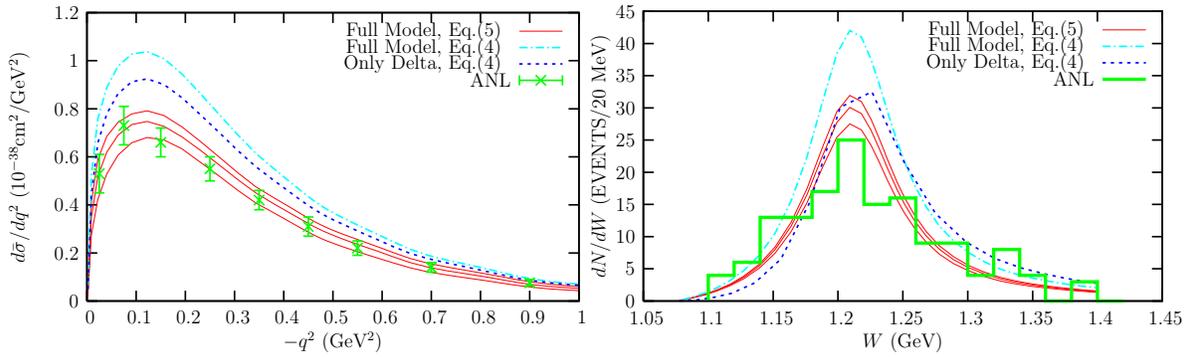}\\\vspace{0.2cm}
\includegraphics[height=.28\textwidth]{dw-anl.epsi}\\\vspace{0.2cm}
\caption{\footnotesize Theoretical results and ANL~\cite{anl} data 
for $\nu_\mu p \to \mu^- p \pi^+$.
  Left: Flux averaged $q^2-$differential cross section
  $\protect\int_{M+m_\pi}^{1.4\,{\rm GeV}}dW
  \frac{d\,\overline{\sigma}_{\nu_\mu \mu^-}}{dq^2dW} $. 
  Right: Flux averaged  $\pi N$ invariant
  mass distribution of events.  Dashed lines stand
  for the contribution of the $\Delta$ pole term with $C_5^A(0)=1.2$ and
  $M_{A\Delta}= 1.05$ GeV.  Dashed--dotted and central solid lines are
  obtained when the full model of Fig.~\ref{fig:diagramas} is
  considered with $C_5^A(0)=1.2,\, M_{A\Delta}= 1.05$ GeV
  (dashed-dotted) and with our best fit parameters $C_5^A(0)=0.867,\,
  M_{A\Delta}= 0.985$ GeV (solid). We also show the 68\% CL bands from
  Eq.~(\protect\ref{eq:besfit}).}\label{fig:anl-bnlq2}
\end{figure}

In Fig.~\ref{fig:anl-bnlq2} (left panel) we present the flux averaged
$q^2$ differential $\nu_\mu p \to \mu^-p\pi^+$ cross sections from the
ANL experiment, which incorporates a $\pi N$ invariant mass cut $W\le
1.4$ GeV.  The agreement with the ANL data is certainly worsened when
the non-resonant terms, required by chiral symmetry, are considered
(dashed-dotted line). This strongly suggests the re-adjustment of the
dominant $C_5^A(q^2)$ form--factor. A $\chi^2-$fit leads to
\begin{equation}
C_5^A(0) = 0.867 \pm 0.075,~ M_{A\Delta}=0.985\pm 0.082\,{\rm
  GeV}, \label{eq:besfit}
\end{equation}
We observe a correction of the order of 30\% to the off diagonal GTR.
Results for total and differential neutrino, antineutrino and NC cross
sections for different isospin channels can be found in
Ref.~\cite{prd}.  In general, the inclusion of the chiral symmetry
background terms brings in an overall improved description of data as
compared to the case where only the $\Delta$ pole mechanism is
considered (see for instance the right panel of
Fig.~\ref{fig:anl-bnlq2}).
\begin{figure}
\includegraphics[height=.28\textwidth]{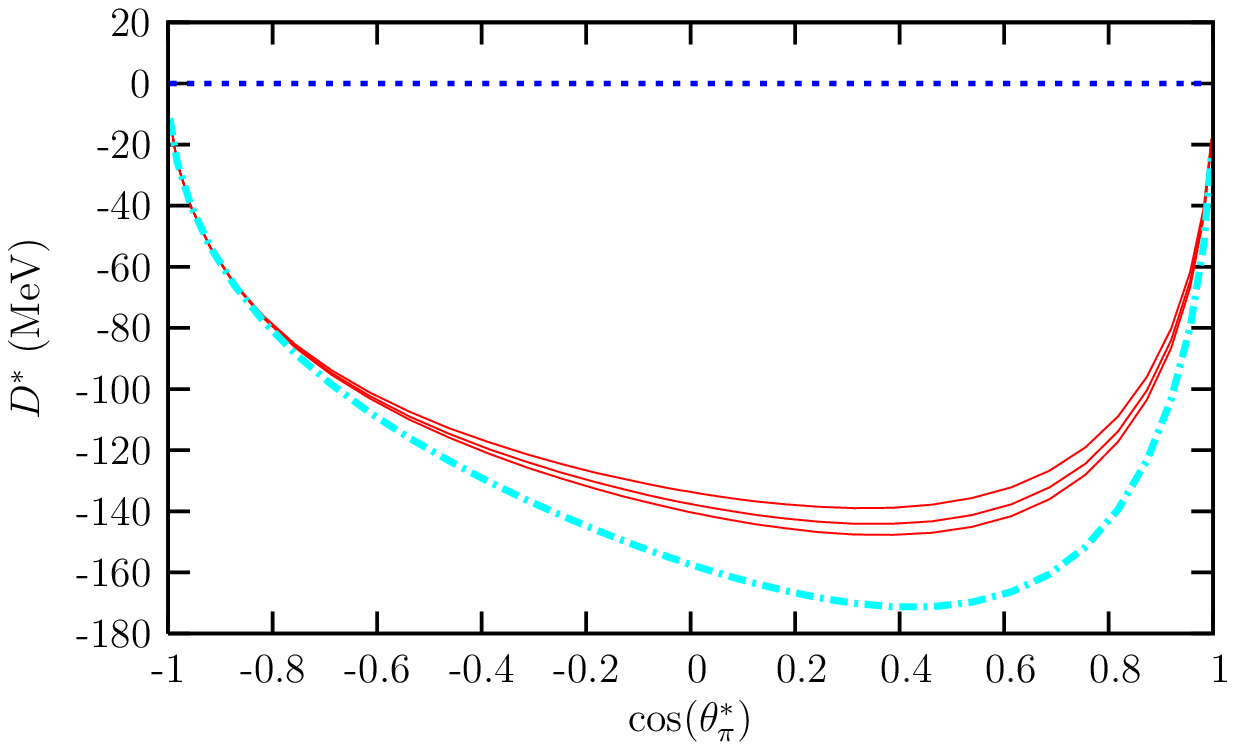}\\\vspace{0.2cm}
\includegraphics[height=.28\textwidth]{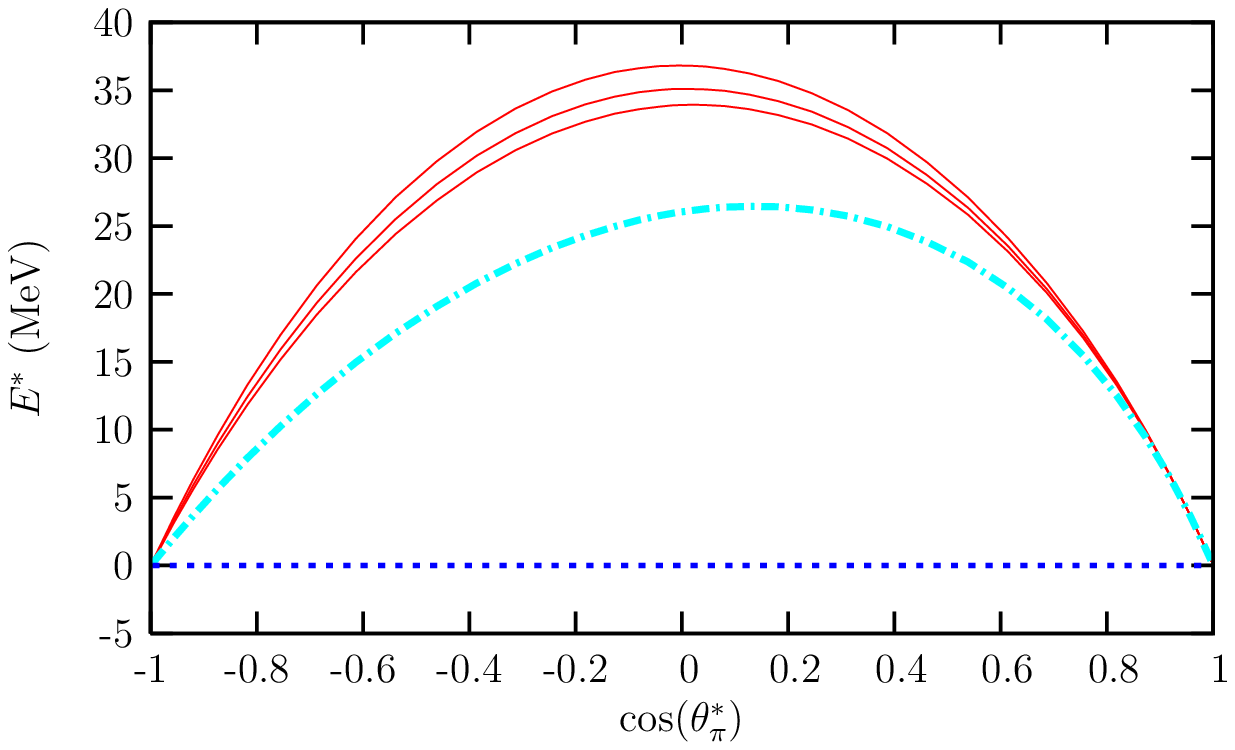}
\vspace{.4cm}
\caption{\footnotesize $D$ and $E$ structure functions
  (Eq.~(\ref{eq:phipi})) for the $\nu_\mu n \to \mu^- p\pi^0$ and
  $\bar\nu_\mu p \to \mu^+ n\pi^0$ reactions, as a function of the
  pion polar angle (in the CM $\pi N$ frame).  The neutrino incoming
  energy is $E=1.5$ GeV, $q^2=-0.5$ GeV$^2$, $W=M_\Delta$ GeV. Dashed
  lines stand for the contribution of the excitation of the
  $\Delta^{0}$ resonance and its subsequent decay with $C_5^A(0)=1.2$
  and $M_{A\Delta}= 1.05$ GeV. Dashed--dotted and  solid lines
  are obtained when the full model of Fig.~\ref{fig:diagramas} is
  considered with $C_5^A(0)=1.2,\, M_{A\Delta}= 1.05$ 
  (dashed-dotted) and with our best fit parameters $C_5^A(0)=0.867,\,
  M_{A\Delta}= 0.985$ GeV (solid).}
\label{fig:parity}
\end{figure}

The interference between the $\Delta$ pole and the background terms
produces parity-violating contributions~\cite{prd} (structure
functions $D$ and $E$ in Eq.~(\ref{eq:phipi})) to
$d^{\,5}\sigma_{\nu_l l}/d\Omega(\hat{k^\prime})dE^\prime
d\Omega(\hat{k}_\pi)$, as shown in Fig.~\ref{fig:parity}. Such
dependences on $\phi_\pi$ might be used to constrain the axial form
factor $C_5^A$. On the other hand, the NC cross section dependence on
$\phi_\pi$ constitutes a potential tool to distinguish $\nu_\tau-$
from $\bar\nu_\tau$, below the $\tau-$production threshold, but above
the pion production one~\cite{tau-nc}.  

We have recently extended the present study to describe two pion
production processes near threshold~\cite{twopion}.
\begin{theacknowledgments}
This work was supported by DGI and FEDER funds, under contracts
  FIS2005-00810, FIS2006-03438, BFM2003-00856 and FPA2004-05616, by
  Junta de Andaluc\'\i a and Junta de Castilla y Le\'on under
  contracts FQM0225, SA104/04 and SA016A07.
\end{theacknowledgments}

\bibliographystyle{aipproc}   

\begin{thebibliography}{9}


\bibitem{ncjuan} J. Nieves, M. Valverde and M.J. Vicente Vacas, 
\emph{Phys. Rev.}  \textbf {C73}  025504 (2006).


\bibitem{ccjuan1} J. Nieves, J.E. Amaro and M. Valverde,
  \emph{Phys. Rev.}  \textbf{
    C70}  055503 (2004); \emph{Erratum-ibid} \textbf{C72}, 019902
  (2005); M.  Valverde, J.E. Amaro and J. Nieves, \emph{Phys. Lett.}
  \textbf {B638} 325 (2006).



\bibitem{LS72} C.H. Llewellyn Smith, \emph{Phys. Rep.}  \textbf{3}  261
  (1972); P. A. Schreiner and F. von Hippel,
  \emph{Phys. Rev. Lett.}  \textbf{30}  339 (1973).

\bibitem{FN} G.L. Fogli and G. Nardulli, \emph{Nucl. Phys.} \textbf {B160}
   116 (1979); \emph{Nucl. Phys.}  \textbf {B165} 162 (1980).

\bibitem{ASV98} L. Alvarez-Ruso, S.K. Singh, M.J. Vicente Vacas,
   \emph{Phys. Rev.} \textbf {C57} 2693 (1998); \emph{Phys. Rev.}
   \textbf {C59} 3386 (1999).

\bibitem{SUL03} T. Sato, D. Uno and T.S.H. Lee, \emph{Phys. Rev.}
    \textbf {C67} 065201 (2003).

\bibitem{Paschos}  E.A. Paschos, J.-Y. Yu and M. Sakuda,  \emph
  {Phys. Rev.}   \textbf {D69} 014013 (2004); 
O. Lalakulich and E.A. Paschos, \emph{Phys. Rev.}  
\textbf{D71} 074003 (2005); O. Lalakulich, E.A. Paschos and G. Piranishvili,
  \emph {Phys. Rev.}  \textbf {D74}  014009 (2006).


\bibitem{anl} G.M. Radecky et al.,  \emph{Phys. Rev.} \textbf
  {D25} 1161 (1982); J. Campbell et al., Phys. Rev. Lett. {\bf 30}
  (1973) 335.

\bibitem{bnlviejo}  T. Kitagaki et al., \emph{Phys. Rev.} \textbf
  {D34} 2554 (1986).

\bibitem{prd} E. Hern\'andez, J. Nieves and M. Valverde,
  \emph {Phys. Rev.}  \textbf {D76}  033005 (2007).

\bibitem{GNO97} A. Gil, J. Nieves and E. Oset, \emph{Nucl. Phys.} \textbf
{A627}  543 (1997).

\bibitem{tau-nc} E. Hern\'andez, J. Nieves and M. Valverde,
    \emph{Phys. Lett.} \textbf {B647} 452 (2007).

\bibitem{twopion} E. Hern\'andez, J. Nieves, S.K. Singh, M. Valverde
  and M.J. Vicente-Vacas,
  \emph{arXiv:0710:3562 [hep.ph]}

\end{thebibliography}

\end{document}